\documentclass[useAMS,usenatbib]{mn2e}
\usepackage{graphicx}
\usepackage[update,prepend]{epstopdf}

\title[Dark neutron stars]{Dark neutron stars}
\author[P. B. Jones]{P. B. Jones\thanks{E-mail:
p.jones1@physics.ox.ac.uk}  \\
University of Oxford, Department of Physics, Denys Wilkinson Building,\\
Keble Road, Oxford OX1 3RH, U.K.}

\begin{document}

\date{}

\pagerange{\pageref{firstpage}--\pageref{lastpage}} \pubyear{}

\maketitle

\label{firstpage}

\begin{abstract}

There is good evidence that electron-positron pair formation is not present in that section of the pulsar open magnetosphere which is the source of coherent radio emission, but the possibility of two-photon pair creation in an outer gap remains.  Calculation of transition rates for this process based on measured whole-surface temperatures, combined with a survey of $\gamma$-ray, X-ray and optical luminosities, expressed per primary beam lepton, shows that few Fermi LAT pulsars have significant outer-gap pair creation.  For radio-loud pulsars with positive polar-cap corotational charge density and an ion-proton plasma there must be an outward flow of electrons from some other part of the magnetosphere to maintain a constant net charge on the star.  In the absence of pair creation, it is likely that this current is the source of GeV $\gamma$-emission observed by the Fermi LAT and its origin is in the region of the outer gap.  With negative polar-cap corotational charge density, the compensating current in the absence of pair creation can consist only of ions or protons.  These neutron stars are likely to be radio-quiet, have no observable $\gamma$-emission, and hence can be described as dark neutron stars.

\end{abstract}


\section{Introduction}

Gamma-ray emission in the $0.1 - 100$ GeV interval is now the most systematically studied part of the pulsar electromagnetic spectrum apart from the coherent radio emission, the results being published in the Second Fermi LAT Catalogue (Abdo et al 2013) containing spectra for 117 pulsars. Anticipation of it has inspired many theoretical papers, mostly directed towards an understanding of the complete X-ray and $\gamma$-ray spectrum generated by synchrotron and curvature radiation, also the emission geometry in terms of the observed relative pulse longitudes of the radio and $\gamma$-emissions (see Takata, Chang \& Shibata 2008; Venter, Harding \& Guillemot 2009; Wang, Takata \& Cheng 2011; Vigano, Torres \& Marti\'{n} 2015a; Pierbattista al 2016  and papers cited therein).  The present paper is less ambitious being addressed to the implications of the Fermi LAT spectra for the broad nature of the magnetosphere in neutron stars with polar-cap ${\bf \Omega}\cdot{\bf B_{s}}$ either positive or negative (${\bf \Omega}$ is the rotation angular velocity and ${\bf B_{s}}$ the polar-cap magnetic flux density).  The implications and conclusions reached follow from the following basis: that neutron stars which are sources of radio emission with the almost universal large negative spectral index are those with polar-cap ${\bf \Omega}\cdot{\bf B_{s}} < 0$ and an ion-proton plasma.  Observational and other evidence for this conclusion has been summarized previously (Jones 2016) and we believe it to be strong.

The theory of $\gamma$-emission at GeV energies was studied before the launch of Fermi LAT.  Earlier observations (Thompson et al 1996) had shown the process to be remarkably efficient, indicating radiation-reaction-limited curvature radiation as the source (see, for example, Romani 1996).  Our phenomenological analysis of the Fermi LAT spectra described in Section 2 although less detailed than that of Vigano et al confirms this.

As in a previous analysis of radio luminosities (Jones 2014), the energy emitted per unit charge at Goldreich-Julian flux densities (Goldreich \& Julian 1969) removes one source of variability from pulsar to pulsar and proves to be an informative parameter.  We adopt a simplified version of the outer-gap model introduced by Cheng, Ho \& Ruderman (1986), following a very early paper by Holloway (1973), and later extended by Cheng, Ruderman \& Zhang (2000).

 Magnetospheric structure in the vicinity of the light cylinder $R_{LC}$ is remote from the complications possible near polar caps and should have some degree of universality with length-scales linear in the rotation period $P$, the only unknown being the angle $\psi$ between the magnetic dipole and rotation axes.  On this basis, the phenomenological analysis of Section 2 confirms that the Fermi LAT spectra of both young and millisecond pulsars (MSP) are consistent with a simple model of radiation-reaction-limited curvature radiation.  Pair production of the order of the Goldreich-Julian density does occur in a small number of neutron stars with relatively high whole-surface temperatures as a consequence of the Breit-Wheeler process ($\gamma\gamma\rightarrow {\rm e}^{+}{\rm e}^{-}$ ; Breit \& Wheeler 1934). This partially screens the outer-gap acceleration field ${\bf E}_{\parallel}$, reducing it to a lesser value $\tilde{\bf E}_{\parallel}$ in such cases. (In this paper, the subscripts parallel and perpendicular refer to the local magnetic flux direction.) This process is described in some detail in Section 3.
Many Fermi LAT catalogue pulsars have no other observed emission and most of the remainder emit otherwise only at radio frequencies.  But a small number of young short-period pulsars, particularly J0534+2200, also have high-luminosity emission at optical and at soft X-ray energies (0.3 to 10 keV).  The emission of photons of the order of 1 TeV has been observed in the case of J0534+2200 (Aliu et al 2011, Ansoldi et al 2016). A survey of no more than orders of magnitude of the the $\gamma$, X-ray and optical luminosities allows one to infer that there is little or no pair creation in all except a very small number of Fermi LAT pulsars.
 
The implications of this modelling are considered first for neutron stars with polar-cap ${\bf \Omega}\cdot{\bf B_{s}} < 0$ and an ion-proton plasma.  The compensating current is of electrons and the  conclusion is that these comprise almost all of the Fermi LAT catalogue.  Electron acceleration in neutron stars with ${\bf \Omega}\cdot{\bf B_{s}} > 0$ has been studied in some detail
(see, for example, Harding \& Muslimov 2001, 2002; Hibschman \& Arons 2001).  Whilst we must anticipate that such stars exist, our belief is that except when very young, they are neither ordinary coherent radio-emitters or gamma-ray emitters in the Fermi LAT energy interval.  We refer to them as dark neutron stars which should be present, possibly with intensities too low for observation, in other parts of the electromagnetic spectrum.  The arguments for these conclusions are given in Section 5.

\section{Curvature radiation}

We assume here the simplest possible variant of the outer-gap model in which there is an interval of electric field ${\bf E}_{\parallel}$ of length $l_{0}$ within the light cylinder radius $R_{LC}$. A notional estimate of this unscreened field can be obtained directly from the values of $B_{s}$ and $P$. It is,
\begin{eqnarray}
E_{\parallel} = \left(\frac{1}{R_{LC}}\right)\left(\frac{\pi R^{2}_{LC}B_{s}}{cP}\right)\left(\frac{R^{3}}{R^{3}_{LC}}\right) = \frac{\pi B_{s}R^{3}}{cPR^{2}_{LC}}
\end{eqnarray}
where $R$ is the neutron-star radius and $B_{s}$ is specifically the inferred polar-cap dipole field.  The radius of curvature adopted is fixed, $\rho = R_{LC}/\sqrt{3}$, the value at the null point ${\bf \Omega}\cdot{\bf B} = 0$ of the last closed field-line in a magnetically aligned neutron star.  The radius of curvature is not a rapidly varying function of position, being in the aligned case $\rho = R_{LC}/3$ at the equator. In Tables 1-3, the length of the finite-$E_{\parallel}$ region is set at the reference value $l^{ref}_{0} = 0.1R_{LC}$.  We are confident that as functions of $B_{s}$ and $P$, the parameters $l_{0}$, $\rho$ and $E_{\parallel}$ will scale as indicated.  The observed cut-off energies $E_{c}$ in the Fermi LAT catalogue are treated as curvature radiation cut-offs for a mono-energetic group of electrons.  The values of the energy that can in principle be gained, ${\rm e}E_{\parallel}l^{ref}_{0}$ are given in column 4 of Tables 1-3.  
Values of the radiation-reaction-limited electron Lorentz factor $\gamma$ obtained directly from $\rho$ and $E_{c}$ are in column 7.  The radiation-reaction-limited  curvature-photon energy released by an electron traversing $l_{0}$ is,
\begin{eqnarray}
E_{rl} = {\rm e}\tilde{E}_{\parallel}l_{0} = \frac{2{\rm e}^{2}\gamma^{4}l_{0}}{3\rho^{2}},
\end{eqnarray}
neglecting any inverse Compton scattering process that could also contribute to energy loss.  Comparison of $L$ and $E_{rl}$ with the energy $\gamma mc^{2}$ derived from column 7 demonstrates the effect of radiation-reaction-limited emission.
All energies are per electron and in units of GeV.  Estimates of the measured $\gamma$-ray luminosity $L$ and the spin-down energy loss $E_{sd}$ per electron are obtained by assuming Goldreich-Julian fluxes for the alignment case $\psi = \pi$, with circular polar caps of radius,
\begin{eqnarray}
u_{0} = \left(\frac{2\pi R^{3}}{c P f(1)}\right)^{1/2},
\end{eqnarray}
Harding \& Muslimov (2001), with $f(1) = 1.368$ and $R = 1.0\times10^{6}$ cm for consistency with the ATNF Pulsar Catalogue (Manchester et al 2005) from which the values of $B_{s}$ have been drawn.

There is no attempt here to account for the shape of individual spectra in the Fermi LAT catalogue which vary significantly from pulsar to pulsar, particularly in the interval $100$ MeV to $1$ GeV.  The mono-energetic group of electrons assumed does not fit the spectral shapes below $E_{c}$ or the values of the parameter $\Gamma$ in the catalogue.  Clearly this is a consequence of our neglect of contributions to the spectrum from synchrotron radiation and inverse Compton scattering.  Also we ignore  lower-energy groups of electrons which must be present in the source region as a consequence of boundary conditions satisfied by the acceleration field or of pair creation. 

Fig. 1 shows the distributions of $L/E_{rl}$ separately for normal young pulsars and for MSP. Two features are obvious. Both distributions have an order-of-magnitude width and contain a small fraction of pulsars with $L \gg E_{rl}$.  The values of $E_{rl}$ are a substantial under-estimate of $L$ but within the limitation imposed by the small numbers concerned, there is no indication that the two distributions are other than identical. Given the several orders of magnitude differences in typical values of $B_{s}$ and $P$ for the two sets, the latter is an indication that the model is soundly based although it is possible that the assumed value of $l_{0}/\rho^{2}$ is too small.  The are a number of explanations for the widths.  These may be, at least in part, a consequence of the unknown spread in magnitude of the angle $\psi$ between the spin and magnetic moment axes, whose value must affect the geometry of the emission region. It is possible that there may be a caustic in one of the variables defining the $\gamma$-ray beam profile.  A wide spread in observed emissivity per unit charge is also a feature of radio emission (see Jones 2014) for which there may be a similar explanation.  Abdo et al (2013; equation 15) assumed a beam correction factor $f_{\Omega} = 1$ corresponding to a fan-beam sweeping out $4\pi$ steradians.  This is appropriate for a catalogue, but the true value may be more nearly $f_{\Omega} \sim 10^{-1}$. The presence in Tables 1-3 of 9 pulsars for which the observed quantities $L \geq E_{sd}$ suggests that these explanations are relevant to both normal young pulsars and MSP and may explain why the values of $E_{rl}$ underestimate $L$.

\begin{figure}
\includegraphics[trim=15mm 60mm 20mm 130mm,clip,width=84mm]{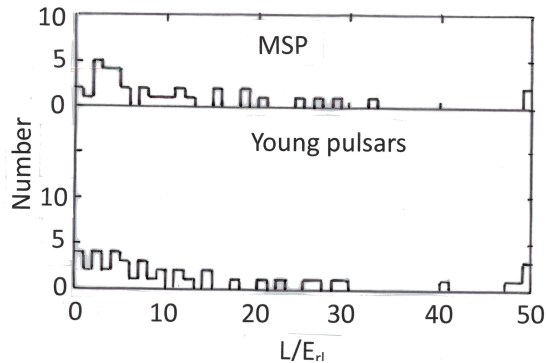}

\caption{These histograms show the distribution of $L/E_{rl}$ separately for 44 young pulsars and for 37 MSP in the Fermi LAT catalogue.  They are small samples but there is no indication that they are drawn from different distributions.  Large values of the ratio are positioned at the extreme right-hand side of each histogram.}
\end{figure}

The remaining comparison to be made in Tables 1-3 is in the ratio of the energy transfer ${\rm e}E_{\parallel}l^{ref}_{0}$, which would occur if the field were the notional unscreened field defined by equation (1), to $E_{rl}$.  Here there are a few very large values, in particular, for the Crab pulsar J0534+2200.  This means that in this case, the notional acceleration field $E_{\parallel}$ very much over-estimates the true field $\tilde{E}_{\parallel}$.
Consideration of possible pair-production processes in regions beyond the null point  ${\bf \Omega}\cdot{\bf B} = 0$ on the last closed flux line shows that the origin of the screening is the Breit-Wheeler process described in this context by Cheng, Ho \& Ruderman (1986).  This ratio given by Tables 1-2 for normal pulsars has some large values for small characteristic age $\tau_{c}$, perhaps the best universal indicator of the whole-surface temperature $T^{\infty}_{s}$ seen by a distant observer.  But for pulsars in Tables 1 and 2 with $\tau_{c} \geq 10^{5}$ yr the mean ratio is $1.70$, almost identical with $1.66$ for the MSP, indicating that our very basic model assumptions are not too badly wrong. This correlation between $E_{\parallel}/\tilde{E}_{\parallel}$ and age is not inconsistent with the presence of the Breit-Wheeler process.  The MSP have characteristic ages of the order of $10^{9}$ yr and in this respect are a homogeneous group, again consistent with middle-aged normal pulsars.  Unfortunately, few whole-surface temperatures are known for Fermi LAT pulsars, but published values will be quoted and used in Section 3.

The conclusion of this Section is that curvature radiation from a Goldreich-Julian flux of electrons is adequate to explain Fermi-LAT $\gamma$-ray luminosities and, specifically, that the observed spectral cut-offs are those of curvature radiation.    The electron acceleration region must be in the open sector of the magnetosphere and on flux lines separate from those which are the source of the coherent radio emission.  Precisely how the electrons arrive there in the absence of self-sustaining Breit-Wheeler pair creation is a question that is considered in Section 5.

\begin{table*}
\caption{The table contains in column 3 the characteristic age, and in column (4) the notional energy transfer in the acceleration gap ${\rm e}E_{\parallel}l^{ref}_{0}$ defined by equation (1) in terms of the period $P$ and the polar-cap dipole field $B_{s}$ given in the ATNF Pulsar Catalogue. The observed spin-down energy loss-rate $E_{sd}$ and luminosity $L$ are those of Abdo et al expressed in GeV per primary electron.  The radiation-reaction-limited values of the Lorentz factor $\gamma$ and consequent energy transfer are contained in columns (7) \& (9) and assume that the acceleration region is set at a reference length $l^{ref}_{0} = 0.1R_{LC}$.  The cut-off energies $E_{c}$ in column 6 are those of Abdo et al.
Column 10 gives the luminosity $L_{X}$ per unit primary electron of X-ray flux in the interval $0.3 - 10$ keV for pulsed X-ray sources. Column 11 gives the luminosity $L_{V}$ of the optical V-band flux, also for pulsed sources.  The data are drawn directly from Tables 15 - 17 of Abdo et al (2013) assuming a beam correction factor $f_{\Omega} = 1$ and are given here for convenient reference.}

\begin{tabular}{@{}lrrrrrrrrrr@{}}
\hline       
Pulsar     &    $P$  &  $\tau_{c}$ &   ${\rm e}E_{\parallel}l^{ref}_{0}$   &   $E_{sd}$   &   $E_{c}$    &       $\gamma$   &    $L$    &    $E_{rl}$  &
$L_{X}$  &  $L_{V}$      \\
\hline
   J-name &  s  &  yr  &  GeV  &  GeV  &  GeV  &      &  GeV  &  GeV &GeV & Gev   \\
\hline
0007+7303 & 0.316 & 1.4E4 & 7.1E4 & 2.6E6 & 4.7 & 5.17E7 & 5.4E5 & 1.4E5 &125& \\   
0106+4855 & 0.083 & 3.1E6 & 1.8E4 & 6.6E5 & 2.7 & 2.75E7 & 4.8E5 & 4.1E4 & &  \\
0205+6449 & 0.066 & 5.4E3 & 5.5E4 & 2.0E7 & 1.6 & 2.14E7 & 1.8E5 & 1.9E4 &6000& \\
0248+6021 & 0.217 & 6.2E4 & 4.9E4 & 1.8E6 & 1.6 & 3.18E7 & 2.1E5 & 2.9E4 & &  \\
0357+3205 & 0.444 & 5.4E5 & 8.1E3 & 3.0E5 & 0.8 & 3.21E7 &   -   & 1.4E4 &$<$41000&  \\
0534+2200 & 0.034 & 1.3E3 & 2.2E6 & 8.3E7 & 4.2 & 2.37E7 & 1.1E5 & 5.6E4 &370000&380  \\
0622+3749 & 0.333 & 2.1E5 & 1.7E4 & 6.3E5 & 0.6 & 2.65E7 &   -   & 8.9E3 & &  \\
0631+1036 & 0.288 & 4.4E4 & 4.4E4 & 1.6E6 & 6.0 & 5.42E7 & 5.2E4 & 1.8E5 & &  \\
0633+0632 & 0.297 & 5.9E4 & 3.6E4 & 1.3E6 & 2.7 & 4.21E7 &   -   & 6.4E4 & &  \\
0633+1746 & 0.237 & 3.4E5 & 1.9E4 & 6.9E5 & 2.2 & 3.65E7 & 6.9E5 & 4.5E5 &81&0.035  \\
0659+1414 & 0.385 & 1.1E5 & 2.1E4 & 7.5E5 & 0.4 & 2.43E7 & 4.8E3 & 5.4E3 &77&0.074 \\
0729-1448 & 0.252 & 3.5E4 & 5.6E4 & 2.1E6 &   - &    -   &   -   &   -   & &  \\
0734-1559 & 0.156 & 2.0E5 & 3.8E4 & 1.4E6 & 3.2 & 3.59E7 &   -   & 6.4E4 & &  \\
0742-2822 & 0.167 & 1.6E5 & 4.0E4 & 1.4E6 & 1.6 & 2.92E7 & 9.3E4 & 2.6E4 & &  \\
0835-4510 & 0.089 & 1.1E4 & 2.8E5 & 1.0E7 & 3.0 & 2.92E7 & 1.3E5 & 4.9E4 &92&0.019 \\
0908-4913 & 0.107 & 1.1E5 & 7.4E4 & 2.7E6 & 0.5 & 1.71E7 & 1.9E5 & 4.0E3 & &  \\
0940-5428 & 0.088 & 4.2E4 & 1.5E5 & 5.3E6 &   - &    -   &   -   &   -   & & \\
1016-5857 & 0.107 & 2.1E4 & 1.7E5 & 6.3E6 & 6.0 & 3.91E7 & 1.3E5 & 1.3E5& &  \\
1019-5749 & 0.162 & 1.2E5 & 4.5E4 & 1.6E6 &   - &    -   &   -   &   -   & &  \\
1023-5746 & 0.111 & 4.6E3 & 3.5E5 & 1.3E7 & 2.5 & 2.95E7 &   -   & 4.1E4 & &  \\
1028-5819 & 0.091 & 9.0E4 & 9.6E4 & 3.5E6 & 4.6 & 3.39E7 & 6.7E5 & 8.7E4 & &  \\
1044-5737 & 0.139 & 4.0E4 & 9.5E4 & 3.5E6 & 2.8 & 3.31E7 &   -   & 5.2E4 & &  \\
1048-5832 & 0.124 & 2.0E4 & 1.5E5 & 5.5E6 & 3.0 & 3.26E7 & 4.8E5 & 5.5E4 & &  \\
1057-5226 & 0.197 & 5.3E5 & 1.8E4 & 6.7E5 & 1.4 & 2.95E7 & 9.6E4 & 2.3E4 &49&  \\
1105-6107 & 0.063 & 6.3E4 & 1.7E5 & 6.2E6 & 1.3 & 1.97E7 & 3.7E5 & 1.4E4 & & \\
1112-6103 & 0.065 & 3.3E4 & 2.3E5 & 8.2E6 & 6.0 & 3.31E7 & 6.5E5 & 1.1E5 & & \\
1119-6127 & 0.408 & 1.6E3 & 1.6E5 & 5.8E6 & 3.2 & 4.95E7 & 1.5E6 & 8.8E4&3100&  \\
1124-5916 & 0.135 & 2.8E3 & 3.6E5 & 1.3E7 & 2.1 & 2.98E7 & 1.9E5 & 3.5E4 &3000&  \\
1135-6055 & 0.115 & 2.3E4 & 1.5E5 & 5.7E6 & 2.4 & 2.95E7 &   -   & 4.0E4 & & \\
1357-6429 & 0.166 & 7.3E3 & 1.9E5 & 6.8E6 & 0.9 & 2.40E7 & 5.5E4 & 1.2E4 &60&  \\
1410-6132 & 0.050 & 2.5E4 & 3.4E5 & 1.2E7 &   - &    -   & 9.8E5 &   -   & & \\
1413-6205 & 0.110 & 6.3E4 & 9.6E4 & 3.5E6 & 4.1 & 3.47E7 &   -   & 7.9E4 & & \\
1418-6058 & 0.111 & 1.0E4 & 2.4E5 & 8.5E6 & 5.5 & 3.84E7 & 1.6E5 & 1.2E5 & & \\
1420-6048 & 0.068 & 1.3E4 & 3.4E5 & 1.2E7 & 1.6 & 2.16E7 & 7.7E5 & 1.9E4 & & \\
1429-5911 & 0.116 & 6.0E4 & 9.3E4 & 3.4E6 & 2.2 & 2.87E7 &   -   & 3.5E4 & & \\
1459-6053 & 0.103 & 6.5E4 & 1.0E5 & 3.7E6 & 2.9 & 3.03E7 &   -   & 4.9E4 & & \\
1509-5850 & 0.089 & 1.5E5 & 7.6E4 & 2.7E6 & 4.6 & 3.36E7 & 5.7E5 & 8.6E4 & & \\
1513-5908 & 0.151 & 1.6E3 & 4.4E5 & 1.6E7 &   - &    -   & 6.5E4 &   -   & & \\
1531-5610 & 0.084 & 9.7E4 & 1.0E5 & 3.7E6 &   - &    -   & 4.1E3 &   -   & & \\
1620-4927 & 0.172 & 2.6E5 & 3.0E4 & 1.1E6 & 2.5 & 3.42E7 &   -   & 4.8E4 & & \\
\hline
\end{tabular}
\end{table*}

\begin{table*}
\caption{Continuation of Table 1.}
\begin{tabular}{@{}lrrrrrrrrrr@{}}
\hline
Pulsar & $P$ &  $\tau_{c}$ & ${\rm e}E_{\parallel}l^{ref}_{0}$ & $E_{sd}$ & $E_{c}$ & $\gamma$ &  
 $L$ & $E_{rl}$  & $L_{X}$ & $L_{V}$  \\
\hline
J-name & s & yr & GeV & GeV & GeV &    & GeV & GeV   &GeV & GeV  \\
\hline 
1648-4611 & 0.165 & 1.1E5 & 4.8E4 & 1.8E6 & 6.0 & 4.51E7 & 1.4E6 & 1.5E5  & &\\
1702-4128 & 0.182 & 5.5E4 & 6.2E4 & 2.3E6 & 0.8 & 2.38E7 & 5.3E5 & 1.1E4  & &\\
1709-4429 & 0.102 & 1.7E4 & 2.0E5 & 7.1E6 & 4.2 & 3.42E7 & 1.8E6 & 8.1E4 &500&  \\
1718-3825 & 0.075 & 8.9E4 & 1.2E5 & 4.5E6 & 1.4 & 2.14E7 & 4.8E5 & 1.7E4  & &\\
1730-3350 & 0.139 & 2.6E4 & 1.2E5 & 4.2E6 & 1.2 & 2.49E7 & 1.3E5 & 1.7E4  & &\\
1732-3131 & 0.197 & 1.1E5 & 4.0E4 & 1.5E6 & 1.9 & 3.26E7 & 8.7E4 & 3.4E4  & &\\
1741-2054 & 0.414 & 3.9E5 & 1.0E4 & 3.8E5 & 0.9 & 3.26E7 & 8.4E4 & 1.6E4  & &\\
1746-3239 & 0.200 & 4.8E5 & 1.9E4 & 7.1E5 & 1.5 & 3.03E7 &   -   & 2.5E4  & &\\
1747-2958 & 0.099 & 2.5E4 & 1.7E5 & 6.1E6 & 1.9 & 2.60E7 & 1.4E6 & 2.8E4  & &\\
1801-2451 & 0.125 & 1.5E4 & 1.7E5 & 6.3E6 & 3.0 & 3.27E7 & 9.7E4 & 5.5E4  & &\\
1803-2149 & 0.106 & 8.6E4 & 8.5E4 & 3.1E6 & 3.6 & 3.29E7 &   -   & 6.6E4  & &\\
1809-2332 & 0.147 & 6.8E4 & 6.9E4 & 2.6E6 & 3.4 & 3.60E7 & 9.8E5 & 6.9E4  & &\\
1813-1246 & 0.048 & 4.3E4 & 2.6E5 & 9.6E4 & 2.6 & 2.26E7 &   -   & 3.3E4  & &\\
1826-1256 & 0.110 & 1.4E4 & 2.0E5 & 7.4E6 & 2.2 & 2.82E7 &   -   & 3.5E4  & &\\
1833-1034 & 0.062 & 4.8E3 & 6.1E5 & 2.3E7 & 0.9 & 1.73E7 & 1.1E5 & 8.7E3  & &\\
1835-1106 & 0.166 & 1.3E5 & 4.5E4 & 1.7E6 &   - &    -   & 5.5E4 &   -    & &\\
1836+5925 & 0.173 & 1.8E6 & 1.1E4 & 4.0E5 & 2.0 & 3.18E7 & 7.4E5 & 3.6E4  & &\\
1838-0537 & 0.146 & 4.9E3 & 2.6E5 & 9.5E6 & 4.1 & 3.82E7 &   -   & 8.8E4  & &\\
1846+0919 & 0.226 & 3.6E5 & 1.9E4 & 7.2E5 & 2.2 & 3.59E7 &   -   & 4.4E4  & &\\
1907+0602 & 0.107 & 1.9E4 & 1.8E5 & 6.5E6 & 2.9 & 3.07E7 & 7.2E5 & 5.0E4  & &\\
1952+3252 & 0.040 & 1.1E5 & 2.0E5 & 7.4E6 & 2.5 & 2.10E7 & 1.3E5 & 2.9E4  & &\\
1954+2836 & 0.093 & 6.9E4 & 1.1E5 & 3.8E6 & 3.3 & 3.06E7 &   -   & 5.7E4  & &\\
1957+5033 & 0.375 & 8.4E5 & 7.7E3 & 2.8E5 & 1.0 & 3.27E7 &   -   & 1.8E4  & &\\
1958+2846 & 0.290 & 2.2E4 & 6.2E4 & 2.2E6 & 2.0 & 3.78E7 &   -   & 4.2E4  & &\\
2021+3651 & 0.104 & 1.7E4 & 1.9E5 & 7.2E6 & 3.0 & 3.07E7 & 1.2E7 & 5.1E4  & &\\
2021+4026 & 0.265 & 7.7E4 & 3.6E4 & 1.4E6 & 2.6 & 4.00E7 & 2.9E6 & 5.8E4  & &\\
2028+3332 & 0.177 & 5.8E5 & 2.0E4 & 7.3E5 & 1.9 & 3.15E7 &   -   & 3.3E4  & &\\
2030+3641 & 0.200 & 4.9E5 & 1.9E4 & 7.0E5 & 1.5 & 3.03E7 & 7.4E5 & 2.5E4  & & \\
2030+4415 & 0.227 & 5.5E5 & 1.6E4 & 5.8E5 & 1.7 & 3.30E7 &   -   & 3.1E4  & &\\
2032+4127 & 0.143 & 1.8E5 & 4.4E4 & 1.6E6 & 3.2 & 3.49E7 & 1.6E6 & 6.2E4  & &\\
2043+2740 & 0.096 & 1.2E6 & 2.5E4 & 9.1E5 & 1.2 & 2.20E7 & 6.3E4 & 1.5E4  & &\\
2055+2539 & 0.320 & 1.2E6 & 7.5E3 & 2.7E5 & 1.1 & 3.20E7 &   -   & 2.0E4  & &\\
2111+4606 & 0.158 & 1.7E4 & 1.3E5 & 4.5E6 & 5.0 & 4.19E7 &   -   & 1.2E5  & &\\
2139+4716 & 0.283 & 2.5E6 & 5.9E3 & 2.2E5 & 1.3 & 3.25E7 &   -   & 2.4E4  & &\\
2229+6114 & 0.052 & 1.0E4 & 5.0E5 & 1.8E7 & 4.3 & 2.75E7 & 1.6E4 & 6.6E4 &340&  \\
2238+5903 & 0.163 & 2.7E4 & 1.0E5 & 3.7E6 & 2.1 & 3.17E7 &   -   & 3.7E4  & &\\
2240+5832 & 0.140 & 1.4E5 & 5.0E4 & 1.8E6 & 3.0 & 3.39E7 & 6.6E5 & 5.7E4  & &\\
                
\hline

\end{tabular}
\end{table*}

\begin{table*}
\caption{Continuation of Table 1 to include the MSP.  Here the period $P$ is in units of milliseconds. There are no pulsed optical sources in this Table.}
\begin{tabular}{@{}lrrrrrrrrr@{}}
\hline                                                              
MSP & $P$ & $\tau_{c}$ & ${\rm e}E_{\parallel}l^{ref}_{0}$ & $E_{sd}$  & $E_{c}$  & $\gamma$  & $L$ &  
$E_{rl}$ & $L_{X}$ \\
\hline
J-name & ms & yr & GeV & Gev & GeV &    & Gev & GeV  &GeV \\
\hline

0023+0923 & 3.05 &   -   & 1.3E4 & 4.8E5 & 1.4 & 7.35E6 & 1.5E4 & 5.8E3  &\\
0030+0451 & 4.86 & 7.6E9 & 6.2E3 & 2.3E5 & 1.8 & 9.33E6 & 3.8E4 & 9.4E3  &160\\
0034-0534 & 1.88 & 6.0E9 & 1.8E4 & 6.8E5 & 1.8 & 6.80E6 & 1.3E4 & 5.0E3  &\\
0101-6422 & 2.57 & 7.9E9 & 1.2E4 & 4.2E5 & 1.5 & 7.10E6 & 1.3E4 & 5.9E3  &\\
0102+4839 & 2.96 &   -   & 1.4E4 & 5.1E5 & 3.2 & 9.58E6 & 2.6E5 & 1.7E4  &\\
0218+4232 & 2.32 & 4.8E8 & 5.2E4 & 1.9E6 & 4.6 & 9.97E6 & 3.0E5 & 2.6E4  &3000\\
0340+4130 & 3.29 &   -   & 8.5E3 & 3.8E5 & 2.6 & 9.26E6 & 3.5E5 & 1.3E4  &\\
0437-4715 & 5.76 & 1.6E9 & 1.1E4 & 4.3E5 & 1.1 & 8.38E6 & 1.8E3 & 5.2E3  &86\\
0610-2100 & 3.86 & 4.9E9 & 9.7E3 & 3.6E5 & 1.6 & 8.31E6 & 4.2E5 & 7.4E3  &\\
0613-0200 & 3.06 & 5.1E9 & 1.2E4 & 4.4E5 & 2.5 & 8.93E6 & 9.8E4 & 1.2E4  &\\
0614-3329 & 3.15 & 2.8E9 & 1.6E4 & 5.7E5 & 3.9 & 1.05E7 & 1.2E6 & 2.3E4  &\\
0751+1807 & 3.48 & 7.1E9 & 9.1E3 & 3.3E5 & 2.6 & 9.44E6 & 1.1E4 & 1.4E4  &\\
1024-0719 & 5.16 & 4.4E9 & 7.7E3 & 2.8E5 &   - &    -   & 3.2E3 &   -    &\\
1124-3653 & 2.41 &   -   & 1.4E4 & 5.2E5 & 2.5 & 8.24E6 & 1.3E5 & 1.1E4  &\\
1125-5825 & 3.10 & 8.1E8 & 3.0E4 & 1.1E6 & 4.8 & 1.11E7 & 9.6E4 & 2.9E4  &\\
1231-1411 & 3.68 & 2.6E9 & 1.4E4 & 5.2E5 & 2.7 & 9.74E6 & 6.9E4 & 1.5E4  &\\
1446-4701 & 2.19 & 3.5E9 & 2.0E4 & 7.5E5 & 3.0 & 8.48E6 & 3.9E4 & 1.4E4  &\\
1514-4946 & 3.59 & 3.0E9 & 1.3E4 & 4.9E5 & 5.3 & 1.21E7 & 1.5E5 & 3.6E4  &\\
1600-3053 & 3.60 & 6.0E9 & 9.5E3 & 3.5E5 & 2.0 & 8.75E6 & 7.4E4 & 9.8E3  &\\
1614-2230 & 3.15 & 5.2E9 & 1.2E4 & 4.2E5 & 1.9 & 8.22E6 & 4.2E4 & 8.7E3  &\\
1658-5324 & 2.44 & 3.5E9 & 1.8E4 & 6.7E5 & 1.4 & 6.82E6 & 6.7E4 & 5.3E3  &\\
1713+0747 & 4.57 & 8.5E9 & 6.3E3 & 2.3E5 & 2.7 & 1.05E7 & 8.5E4 & 1.6E4  &\\
1741+1351 & 3.75 & 2.0E9 & 1.6E4 & 5.9E5 &   - &    -   & 7.7E3 &   -    &\\
1744-1134 & 4.07 & 7.2E9 & 7.6E3 & 2.8E5 & 1.2 & 7.69E6 & 3.7E4 & 5.2E3  &\\
1747-4036 & 1.65 & 2.0E9 & 3.6E4 & 1.4E6 & 5.4 & 9.39E6 & 1.6E5 & 2.8E4  &\\
1810+1744 & 1.66 &   -   & 2.1E4 & 7.7E5 & 3.2 & 7.90E6 & 2.2E5 & 1.4E4  &\\
1823-3021A& 5.44 & 2.5E7 & 9.6E4 & 3.5E6 & 2.5 & 1.08E7 & 3.0E5 & 1.5E4  &\\
1858-2216 & 2.38 &   -   & 1.1E4 & 4.1E5 & 1.7 & 7.22E6 & 2.9E4 & 6.9E3  &\\
1902-5105 & 1.74 & 3.0E9 & 2.8E4 & 1.0E6 & 3.4 & 8.19E6 & 5.3E4 & 1.6E4  &\\
1939+2134 & 1.56 & 2.3E8 & 1.1E5 & 4.1E6 &   - &    -   & 5.2E4 &   -    &2300\\
1959+2048 & 1.61 & 1.5e9 & 4.2E4 & 1.5E6 & 1.4 & 5.94E6 & 1.3E5 & 4.6E3  &\\
2017+0603 & 2.90 & 5.5e9 & 1.2E4 & 4.3E5 & 3.4 & 9.71E6 & 3.3E5 & 1.8E4  &\\
2043+1711 & 2.38 & 7.2E9 & 1.3E4 & 4.7E5 & 3.3 & 9.00E6 & 3.1E5 & 1.7E4  &\\
2047+1053 & 4.29 &   -   & 1.1E4 & 4.0E5 & 2.0 & 9.27E6 & 9.7E4 & 1.0E4  &\\
2051-0827 & 4.51 & 5.6E9 & 7.8E3 & 2.9E5 & 1.3 & 8.17E6 & 2.1E4 & 5.9E3  &\\
2124-3358 & 4.93 & 3.8E9 & 8.7E3 & 3.2E5 & 1.6 & 9.02E6 & 1.9E4 & 8.1E3  &34\\
2214+3000 & 3.12 & 3.5e9 & 1.4E4 & 5.2E5 & 2.2 & 8.61E6 & 2.7E5 & 1.1E4  &\\
2215+5135 & 2.61 & 1.2E9 & 2.9E4 & 1.1E6 & 3.4 & 9.38E6 & 1.8E5 & 1.8E4  &\\
2241-5236 & 2.19 & 5.2E9 & 1.7E4 & 6.1E5 & 3.0 & 8.48E6 & 2.6E4 & 1.4E4  &\\
2302+4442 & 5.19 & 6.2E9 & 6.5E3 & 2.4E5 & 2.1 & 1.00E7 & 3.9E5 & 1.2E4  &\\

\hline
\end{tabular}

\end{table*}

\section{Screening by pair creation}

There are a limited number of processes to be considered in relation to screening by electron-positron pair creation.  We refer to Daugherty \& Harding (1983) for an extensive survey of single-photon magnetic conversion.  The asymptotic expression for the transition rate given by equation (17) of their paper has to be evaluated at the null point ${\bf \Omega}\cdot{\bf B} = 0$ on the last closed flux line.  Let us assume that an attenuation rate of $10^{-10}$ cm$^{-1}$ can be regarded as negligible.  Then for $5$ GeV photons with $k_{\perp}/k = 0.1$ this rate requires a local magnetic flux density $B \approx 10^{-4}B_{c} =4.4\times 10^{9}$ G.  For the Crab pulsar, this would be at a null point radius $r \approx 10R$, which is possible only for a very small interval of $\psi$ near $\pi/2$. Thus there could be some contribution from this process but it is unlikely to be significant.  Inverse Compton scattering of black-body photons by electrons with the Lorentz factors given in Tables 1-3 is in principle a source of photons energetic enough for magnetic conversion, but the cross-section is small at the Lorentz-invariant total energy squared $s \approx 2mc^{2}\gamma \epsilon_{bb} \approx 10^{4} (mc^{2})^{2}$ of the present case.  (Here $m$ is the electron mass and $\epsilon_{bb} \sim k_{B}T^{\infty}_{s}$ is the photon energy in the fixed observer frame for whole-surface temperature $T^{\infty}_{s}$.) Following Cheng, Ho \& Ruderman, the important process capable in principle of giving a stable and significant rate of pair production is the collision of black-body and curvature radiation photons.  The Lorentz-invariant total energy squared is $s = 4\epsilon \epsilon_{bb}\sin^{2}(\zeta/2) > 4(mc^{2})^{2}$, where $\zeta$ is the angle between the momenta of the photons in the fixed observer frame.  The cross-section in the centre-of-momentum frame is,
\begin{eqnarray}
\sigma^{c} = \frac{3\sigma_{T}}{16}\left(1 - w^{2}\right)\left((3 - w^{4})\ln\frac{1 + w}{1 - w} - 2w(2 - w^{2})\right),
\end{eqnarray}
in terms of the Thomson cross-section $\sigma_{T}$, with $w^{2} = (s - 4m^{2})/s$.  The transition rate per unit volume is Lorentz invariant (see Breit \& Wheeler 1934) and in consequence we can write down directly in the fixed observer frame an expression for the number of pairs produced per primary electron traversing the emission region.  It is,
\begin{eqnarray}
n_{\pm} = \frac{1}{2}\left(\frac{l_{0}}{c}\right)^{2}\int^{\infty}_{0}d\epsilon
\int^{\infty}_{0}d\epsilon_{bb}F_{c}(\epsilon)F_{bb}(\epsilon_{bb})\sigma^{c}(s)\sin^{2}\frac{\zeta}{2},
\end{eqnarray}
in which $F_{c}$ is the rate of curvature radiation photon creation per electron, and $F_{bb}$  the blackbody flux in the fixed observer frame.  Equation (5) assumes, for simplicity, that acceleration to the radiation-reaction-limited Lorentz factor is rapid so that curvature radiation emission with cut-off $E_{c}$ occurs only within the emission region of length $l_{0}$.  The blackbody flux at radius $r$ is,
\begin{eqnarray}
F_{bb} = \frac{1}{4\pi^{2}}\frac{\epsilon^{2}_{bb}}{\hbar^{3}c^{2}}\frac{R^{2}}{r^{2}}n(\epsilon_{bb})
\end{eqnarray}
in which $n$ is here the blackbody occupation number. (Any complications arising from the composition of the atmosphere are ignored.)
We obtain $F_{c}$ from the curvature radiation power loss for an electron of fixed $\gamma$, as is appropriate in the radiation-reaction limit.  It is,
\begin{eqnarray}
F_{c} = \frac{\sqrt{3}}{2\pi}\left(\frac{2}{3}\right)^{1/3}\frac{{\rm e}^{2}}{\hbar^{2} c}\left(\frac{\hbar c}{\rho E_{c}}\right)^{2/3} \int^{\infty}_{\epsilon/E_{c}}K_{5/3}(x)dx,
\end{eqnarray}
where $K$ is a modified Bessel function of the second kind.  The number of pairs given by equation (5) is proportional to $P^{-2/3}$, given that $\rho = R_{LC}/\sqrt{3}$.

The length interval $l_{0}$ in which the radiation-reaction-limited emission occurs is broadly tangential to the light cylinder at $r \approx R_{LC}$ owing, in part, to the Lorentz transformation from the rotating to the fixed observer frame in which equation (5) is evaluated.  Aberration of the black-body photons is negligible.  Thus we assume a reference value $\zeta = \pi/2$.  The evaluation of equation (5) for $n^{ref}_{\pm}$ shown in Fig. 2 also assumes a reference period $P^{ref} = 0.1$ s, $l_{0} = l^{ref}_{0}$, radius $r = R_{LC}$ and neglects any variation of $r$ over the acceleration interval.  The number of pairs in a general case is then,
\begin{eqnarray}
n_{\pm} = n^{ref}_{\pm}\left(\frac{P^{ref}}{P}\right)^{2/3}\left(\frac{R_{LC}}{r}\right)^{2}\left(\frac{l_{0}}{l^{ref}_{0}}\right)^{2}\left(\frac{\sin^{2}\zeta/2}{\sin^{2}\pi/4}\right).
\end{eqnarray}
The reference value $n^{ref}_{\pm}$ is shown in Fig. 2 as a function of $T^{\infty}_{s}$ for  values of $E_{c}$ relevant to the Fermi LAT catalogue.  This refers only to pair creation inside the acceleration region.  Pair creation beyond this is also present, the number being $n^{sec}_{\pm}$ per primary electron, and undoubtedly makes a contribution to synchrotron radiation at X-ray energies, but is not considered here in relation to screening.

The connection between $n_{\pm}$ and the level of screening can be established only by detailed model calculation.  Our assumption here is simply that $n_{\pm} \approx 1$ represents a large reduction in $E_{\parallel}$ or in $l_{0}$ or both. The most recent compilation of pulsar black-body temperatures is that of \"{O}zel (2013). Unfortunately, surface temperature is one of the more interesting but elusive pulsar parameters and of those listed, only J0633+1746 ($6.3\times 10^{5}$ K), J0659+1414 ($5.7\times 10^{5}$ K), J0835-4510 ($5.8\times 10^{5}$ K), J1057-5226 ($5.9\times 10^{5}$ K), J1119-6127 ($6.2\times 10^{5}$ K) and J1709-4429 ($5.8\times 10^{5}$ K) also appear in the Fermi LAT catalogue. Values of $n_{\pm}$ obtained from these temperatures are, for J0835-4510 ($4\times 10^{-2}$), J1057-5226 ($1.3\times 10^{-2}$) and J1709-4429 ($7\times 10^{-2}$). The very young pulsar J1119-6127 has $n_{\pm} = 0.55$, indicative of partial screening. The remaining two have values less than $10^{-2}$. The case of J0534+2200, the Crab pulsar, is different.  Yakovlev \& Pethick (2004) give only an upper limit of $T^{\infty}_{s} < 2.0\times 10^{6}$ K leading to $n_{\pm} = 4.1$, a high level of screening consistent with the large ratio of ${\rm e}E_{\parallel}l^{ref}_{0}$ to either $L$ or $E_{rl}$ given in Table 1. With regard to the MSP in Table 3, PSR J0218+4232 has a large value of $E_{\parallel}$.  Even this case, to have $n_{\pm} = 1$ would require $n^{ref}_{\pm} = 0.08$ and $T^{\infty}_{s} = 6.5\times 10^{5}$, an unlikely temperature. 

The above $n_{\pm}$ are for the reference values of $\zeta$, $l_{0}$ and for $r=R_{LC}$. It might appear that they would be much increased for $r \ll R_{LC}$.  But this is not the case because the factor,
\begin{eqnarray}
\left(\frac{R_{LC}}{r}\right)^{2}\left(\frac{\sin^{2}\zeta/2}{\sin^{2}\pi/4}\right) \approx \frac{R_{LC}}{8r},
\end{eqnarray}
evaluated for a dipole field at $0.1R_{LC} < r < R_{LC}$, is of order unity.  Of course, pair creation numbers large compared with unity in the acceleration region are not possible owing to charge separation and the formation of charge densities large compared with Goldreich-Julian values.  The number of outward moving pairs formed in the field-free region beyond $l_{0}$ is likely to be not more than of the order of $n^{sec}_{\pm} < 10$ for the reasons stated below, although inward moving pairs will be more numerous in this case because $\zeta \approx \pi$. But these latter pairs will be incident on the polar cap, though the curvature photons from which they are produced will mostly pass by the neutron star and may be observable, as noted by Takata, Chang \& Shibata (2008).  Heating of the polar cap by electrons or positrons accelerated inward would be observable if the flux were approaching Goldreich-Julian values because even after curvature radiation emission, the remaining energy at the polar cap would be of the order of $500$ GeV.

\begin{figure}
\includegraphics[trim=5mm 15mm 5mm 35mm,clip,width=84mm]{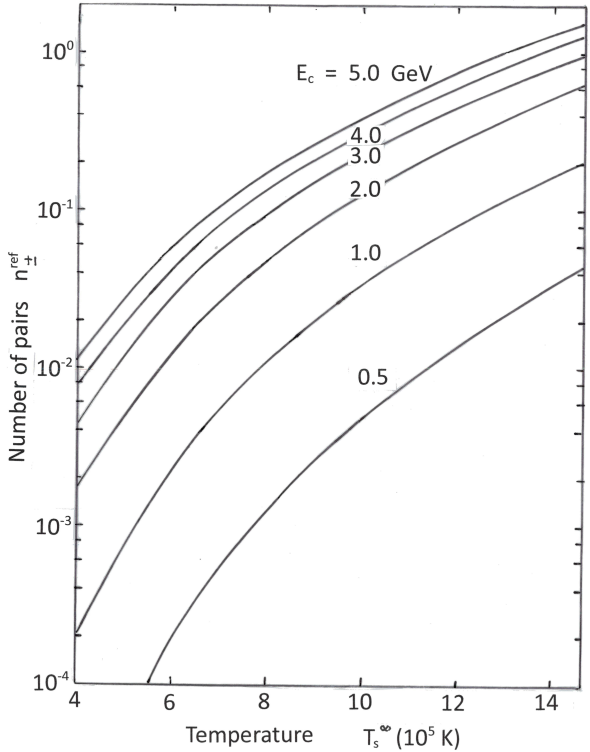} 

\caption{The number of pairs created per primary electron is $n^{ref}_{\pm}$ obtained by evaluating equation (5) for the reference values $P^{ref} = 0.1$ s, $l_{0}^{ref} = 0.1R_{LC}$, $r = R_{LC}$ and $\zeta = \pi/2$, as a function of the neutron-star blackbody temperature $T^{\infty}_{s}$ and curvature-radiation critical energy $E_{c}$.  The radius of curvature is $\rho = R_{LC}/\sqrt{3}$.  Values of $n_{\pm}$ for a particular pulsar are obtained by scaling from the reference values as in equation (8).}

\end{figure} 

Dense pair production, $n^{sec}_{\pm} \gg 1$ is possible above the polar caps of ${\bf \Omega}\cdot{\bf B}_{s} > 0$ neutron stars and has been comprehensively studied (see, for example, Harding \& Muslimov 2001, 2002). It is possible only because the transition rate for the process concerned, single-photon magnetic conversion, is a rapidly increasing function of $k_{\perp}B$, the photon perpendicular momentum component and local magnetic field.  Thus the transition rate is negligible at small $k_{\perp}$ within the finite-field region needed for acceleration but owing to flux-line curvature and possibly aberration, becomes significant in the field-free region above it.  The acceleration region may be limited in extent but is not completely screened. This condition is met only at low altitudes above the polar cap and is not realized elsewhere even in the Crab pulsar.  For photon-photon interaction, the ratio of $n^{sec}_{\pm}$ to $n_{\pm}$ is roughly equal to the effective track length outside the acceleration region divided by $l_{0}$.

A further source of Breit-Wheeler pairs is the interaction between curvature photons in a beam of electrons traversing a curved orbit, giving a rate independent of neutron-star temperature  This has been evaluated approximately for the Crab pulsar and $n_{\pm}$ is several orders of magnitude smaller than unity.  It is $\propto B_{s} P^{-7/3}\langle \sin^{10/3}(\zeta/2)\rangle$ and so is generally negligible for both normal pulsars and MSP.  An average value $\zeta = 0.2$ was assumed for the angle arising from aberration and flux-line curvature, reducing the rate by several orders of magnitude.

\section{Optical and soft X-ray emission}

A small minority of pulsars also have observable optical and soft X-ray emission (see Abdo et al 2013). The most outstanding case is that of J0534+2200.  The cut-off at $E_{c}= 4.2$ GeV appears well-established and, with the scaled value of $\rho$, indicates that a large fraction of the electrons have a  Lorentz factor $\sim 2\times 10^{7}$.  Its luminosities can be compared with those of other Fermi LAT pulsars by
reference to the values of $L$, $L_{X}$ and $L_{V}$ listed in Tables 1-3 and drawn directly from Abdo et al.  It is obvious that the Crab pulsar is truly {\it sui generis}. The ratio $L/L_{X} = 0.30$, typically three orders of magnitude smaller than for any other Fermi LAT pulsar, for example, $1410$ for the Vela pulsar.  This includes the MSP whose distribution of $L_{X}$, though wide, does not obviously differ from the young pulsars of Tables 1 and 2.  We have regarded the primary electrons as homogeneous in energy but in reality there must be some low-energy components if there is a lateral distribution of $E_{\parallel}$.  Considered with the transition rates calculated in Section 3, these luminosities are consistent with there being little or no pair creation in outer gaps except for J0534+2200 and J1119-6127.  Certainly the luminosity ratios for J0534+2200 represent a break from the remaining young pulsars.  This is also indicated by the MSP listed with $L_{X}$ values, whose periods, with the exception of J1939+2134, are longer than average.

The origin of the electrons assumed in the acceleration region was not considered in Section 2, but is it possible that they could be a result of self-sustaining two-photon pair creation?  The electron and positron in a pair can be seen as roughly equivalent to a single electron track length $l_{0}$, except that they have small initial Lorentz factors and non-zero pitch angles, of order $\alpha \sim 0.1$, arising from flux-line curvature.  They emit both synchrotron and curvature radiation with a critical energy,
\begin{eqnarray}
E^{sync}_{c} = \frac{3\hbar c \gamma^{3}}{2\rho}\left(1 + 3\xi + \xi^{2}\right)^{1/2},
\end{eqnarray}
in which
\begin{eqnarray}
\xi = \frac{e B(r)\rho \sin\alpha}{\gamma m c^{2} \cos^{2}\alpha}
\end{eqnarray}
represents the effect of synchrotron radiation (we refer to Vigano et al 2015b, and to Chang \& Zhang 1996).  The $E_{c}$ values obtained by Abdo et al have been used to find the values of $\gamma$ in Tables 1-3. Let us label these as $\gamma = \gamma_{0}$.  Then it follows by direct inspection of equations (10) and (11) that $E^{sync}_{c}(\gamma) < E^{sync}_{c}(\gamma_{0})$ for all $\gamma < \gamma_{0}$ so that the threshold for two-photon pair creation is not reached by photons emitted during the early acceleration interval of electron or positron.  Thus a reasonable approximate criterion for self-sustaining pair creation would be $n_{\pm} > 1$.  Therefore
self-sustaining pair creation appears possible in J0534+2200 and J1119-6127, but not otherwise or in any of the MSP, leading to the conclusion that there must be some other source of outward-accelerated electrons.  This is further considered in the following Section.

For either inverse Compton scattering or synchrotron radiation sources the J0534+2200 flux ratios require both a large flux of relatively low-energy electrons and positrons, and photon fluxes that can be supplied only by coherent radio-frequency emission (Cheng \& Ruderman 1977).  this is consistent with our conclusion that J0534+2200 is probably the only example of self-sustaining two-photon pair creation. We refer to Lominadze, Machabeli \& Usov (1983), Malov \& Machabeli (2001), and Petrova (2003) for recent models for the Crab pulsar optical emission.

Owing to the complex pattern of the radio emission (see, for example, Moffet \& Hankins 1996) the sign of ${\bf \Omega}\cdot{\bf B}_{s}$ above the polar cap in the Crab pulsar is less obvious than in other cases. The radio-frequency profile has two peaks at longitudes essentially identical with those at optical and X-ray frequencies which is possible only with a source near the light cylinder.  But there is minor emission
described as the low-frequency component (LFC) preceding the main component by 40 deg in phase, almost the same difference as is present in the Vela pulsar between its single well-established radio profile and the optical profile main pulse. It also has the large negative spectral index almost universal in radio pulsars.  A high-resolution polarization study at 1.4 GHz (Slowikowska et al 2015) has found that circular polarization is weak or non-existent in the Crab profile except in the LFC in which it is clear, strong and a slowly varying function of longitude. These authors also propose that its source is at low altitude.  Such circular polarization  is an unambiguous indicator of emission from an ion-proton plasma (Jones 2016) and is consistent with their conclusion.

\section{Implications and conclusions}

There must be a net electric charge on a neutron star which is constant when averaged over a rotation period or longer.  In the ${\bf \Omega}\cdot{\bf B}_{s} < 0$ case, an ion-proton plasma is accelerated outwards on the sub-set of magnetic flux lines in the open sector which do not pass through the ${\bf \Omega}\cdot{\bf B} = 0$ null surface.  This plasma is believed to be the source of the coherent radio emission (see Jones 2016).  Although it is not impossible that a fraction of this plasma should re-enter the magnetosphere at the light cylinder, most of it will move outwards into space, so building up a net negative charge on the star. (In the ${\bf \Omega}\cdot{\bf B}_{s}  > 0$ case, the outward flux is of electrons and the increasing net charge would be positive.)  Two outcomes would be possible if the currents were not balanced: the flow of ion-proton plasma and the coherent emission would cease or the structure of the outer magnetosphere would change so as to allow electron emission, though not necessarily in a time-independent way.  The likely region of change would be that beyond the ${\bf \Omega}\cdot{\bf B} = 0$ surface where the change of corotational charge density facilitates the outward acceleration of electrons.

Pair creation and magnetospheric structure are connected because an outward flow of secondary pairs can, if the density is high and the velocities not too close to $c$, change its mean electron-positron relative velocity so as to maintain a Goldreich-Julian charge density, screening any $E_{\parallel}$-component that may have been present.  The necessary balancing current arises if a small fraction of either low-energy positrons or electrons is retained in the magnetosphere.

However, most radio pulsars have periods and fields that do not support pair creation unless flux-line curvature is much smaller than  dipole (see Harding \& Muslimov 2002).  The extreme example is J2144-3933 whose period, determined by Young, Manchester \& Johnston (1999) is $8.51$ s, prompting these authors to question whether some some source of radio emission other than pair plasma might exist. (If included in Table 2, its entry in column 4 would be just 18 GeV.) This pulsar is consistent only with an ion-proton plasma source.  It might be argued that neutron stars unable to support polar-cap pair creation maintain net charge constancy through pair creation elsewhere. However, the relative values of $L$, $L_{X}$ and $L_{V}$ for J0534+2200 and others in Tables 1-3,  and the transition rates in Fig. 2, both show that there can be little or no pair creation in the outer gap of all except a very small number of Fermi LAT pulsars.  Consequently, the model of a corotational magnetosphere must be modified to allow electrons to enter the acceleration region beyond the ${\bf \Omega}\cdot{\bf B} = 0$ null surface.

A modification of this kind has been discussed recently by Melrose \& Yuen (2012, 2014) who investigated the effect of partial screening of the induction field present in a non-aligned neutron star.  The immediate effect is a departure from corotation so that current flows laterally through the surface separating open and closed sectors.  Thus electrons enter the open region beyond the null surface so providing, in principle, the flux assumed in Section 2.  We refer to Jones (2015) for a brief commentary on this problem.

The case of polar-cap ${\bf \Omega}\cdot{\bf B}_{s} > 0$ neutron stars differs from the above. Curvature radiation pair creation is certainly possible at low altitudes above the polar cap in young pulsars.  But owing to the screening capacity of a high flux of secondary pairs, it is then not possible for an outer gap to form and produce the observed GeV $\gamma$-emission.  A high flux of secondary pairs can always adjust to the local corotational charge density and  at the light cylinder, maintain net charge constancy for the whole neutron star. 

If pair creation were absent, primary electrons would be accelerated outwards on flux lines not passing through the null surface and there would be a balancing current of positive charges entering the open sector beyond the null surface.  The outer region of the closed magnetosphere contains protons or ions in an electrostatic and gravitational equilibrium that precludes positrons, of which there also no sources.  Thus the balancing current must consist of baryonic-mass particles and in consequence, would produce no GeV $\gamma$-emission.  Lorentz factors would be too large  and number densities too small to permit growth of Langmuir modes so that there is no obvious source of coherent radio emission.  The only emission would be curvature radiation (X-rays and $\gamma$-rays) from the primary electrons accelerated from the polar cap.

It is likely that few, if any, of the Fermi LAT pulsars fall into this category.  Abdo et al suggest that the difference in beam correction factors for normal young pulsars between radio and $\gamma$-emission is an adequate explanation for the observed division between radio-loud and radio-quiet.  This is also consistent with the almost complete absence of radio-quiet MSP. (The radio beam correction factor is $\propto P^{-1/2}$.)

Neutron stars with polar-cap ${\bf \Omega}\cdot{\bf B}_{s} > 0$ are presumably formed but we believe that in general they are not radio-loud, at least with the universal large negative spectral index, and are unlikely to be present in the Fermi LAT catalogue. A number of known objects may fall into this category, possibly the Central Compact Objects (CCO; Gotthelf, Halpern \& Alford 2013) and the X-ray Isolated Neutron Stars (XINS; Kaplan \& van Kerkwijk 2009). These, particularly the XINS, are near us and some have observable thermal or magnetospheric X-ray emission enabling measurement of their periods and period derivatives.  It remains possible that they may include ${\bf \Omega}\cdot{\bf B} < 0$ pulsars whose radio beams miss the Earth.

But ${\bf \Omega}\cdot{\bf B}_{s} > 0$ neutron stars, although loosely described as dark, remain interesting because a fraction of their spin-down energy loss may appear in other parts of the electromagnetic spectrum.  The Crab pulsar has been tentatively assigned to the ${\bf \Omega}\cdot{\bf B}_{s} < 0$ class on the basis that the LFC, having circular polarization slowly varying with longitude, is the polar-cap emission. The source of the main and inter-pulse radio components can be only those electrons and positrons that are also the origin of the optical and X-ray emission whose profiles are essentially the same.  Thus coherent emission must be possible at the light cylinder given sufficient electron-positron densities.  (The absence in present observations of such radio components in Vela is a further indication that its secondary pair density at the light cylinder is small or negligible.)  Coherent emission, apart from maser action, depends on fluctuations in local charge density and thus on the local plasma frequency.
This is a function of the particle mass and would indicate polar-cap emission $10 - 100$ GHz, rather than the frequencies usually observed, and with luminosities smaller than observed in the ion-proton case(Jones 2014).  The high-frequency emission of some magnetars (Camilo et al 2007; Serylak et al 2009) may also be an electron-positron phenomenon.  The conclusion is that ${\bf \Omega}\cdot{\bf B}_{s} > 0$ neutron stars may have interesting emission though its observation probably requires more work at higher frequencies and the higher sensitivity of the Square Kilometre Array.

\bsp

\label{lastpage}

\end{document}